\documentclass[english]{article}
\usepackage[affil-it]{authblk}
\usepackage[T1]{fontenc}
\usepackage[latin9]{inputenc}
\usepackage{refstyle}
\usepackage{color,hyperref}
    \catcode`\_=11\relax
    \newcommand\email[1]{\_email #1\q_nil}
    \def\_email#1@#2\q_nil{%
      \href{mailto:#1@#2}{{\emailfont #1\emailampersat #2}}
    }
    \newcommand\emailfont{\sffamily}
    \newcommand\emailampersat{{\color{red}\small@}}
    \catcode`\_=8\relax

\makeatletter

\RS@ifundefined{subref}
  {\def\RSsubtxt{section~}\newref{sub}{name = \RSsubtxt}}
  {}
\RS@ifundefined{thmref}
  {\def\RSthmtxt{theorem~}\newref{thm}{name = \RSthmtxt}}
  
  {}
\RS@ifundefined{lemref}
  {\def\RSlemtxt{lemma~}\newref{lem}{name = \RSlemtxt}}
  {}

\usepackage{amsmath}

\makeatother

\usepackage{babel}

\begin{document}
\title{CLASSICAL DYNAMICS FROM A UNITARY REPRESENTATION OF THE GALILEI GROUP }
\author{A. D. Berm\'udez Manjarres}
\affil{\footnotesize Universidad Distrital Francisco Jose de Caldas\\ Cra 7 No. 40B-53, Bogot\'a, Colombia\\ \email{ad.bermudez168@uniandes.edu.co}}
\author{M. Nowakowski}
\affil{\footnotesize Departamento de F\'isica\\
Universidad de los Andes\\
Cra. 1E No. 18A-10
Bogot\'a, Colombia\\ \email{mnowakos@uniandes.edu.co}}
\author{D. Batic}
\affil{\footnotesize Department of Mathematics\\ Khalifa University of Science and
Technology, Main Campus, Abu Dhabi, United Arab Emirate \\ \email{davide.batic@ku.ac.ae}}

\maketitle
\begin{abstract}
We give a formulation of classical mechanics in the language of operators
acting on a Hilbert space. The formulation given comes from a unitary
irreducible representation of the Galilei group that is compatible
with the basic postulates of classical mechanics, particularly the
absence of an uncertainty principle between the position and momentum
of a particle. It is shown that the theory exposed in the article
contains the Koopman-von Neumann formalism of classical mechanics
as a particular case. 
\end{abstract}

\section{Introduction}

The study of the realizations of the Galilei group in classical mechanics
is different compared to the one given to quantum mechanics. The disparity
in the approaches is due to the differences in the usual mathematical
structures used for both theories. The Hilbert space formalism of
quantum mechanics leads naturally to a study of the unitary representations
of the Galilei group. There is a large literature on this subject,
we will cite the classic works \cite{quantum1,quantum 2,quantum 3,levy quantum,levy,quantum5}
and the more recent ones \cite{quantum 7,quantum 8}. On the other
hand, classical mechanics have been derived from the structure of
the Galilei group in the context of Lagrangian mechanics \cite{levy,landau mechanics}
and as a canonical representation in terms of Poisson brackets in
Hamiltonian mechanics \cite{canonical}. 

However, to study the interplay between quantum and classical mechanics
is desirable to express both theories in mathematical formalism that
are similar to each other. In this regard, there are two ways to proceed.
One way is the rewrite quantum mechanics in the same language of classical
mechanics as it is done in the versions of the theory known as quantum
mechanics in phase space \cite{QMphase1,QMphase2} and geometrical
quantum mechanics \cite{geometric,geometric2}. Conversely, there
is an old, but perhaps not widely known, mathematical formalism due
to of Koopman and von Neumann that puts classical Hamiltonian mechanics
in the same mathematical language of quantum mechanics, i.e., as a
theory of operators acting on a Hilbert space and a statistical interpretation
given by a Born rule \cite{cla1,cla 2} (Ref. \cite{cla 3} gives
a good review on the topic).

The Koopman-von Neumann formalism (hereafter abbreviated as KvN) has
been used to investigate the differences and similarities between classical and quantum mechanics and the overall relation between
both theories. The classical limit of quantum mechanics in phase space
is discussed in \cite{cla 5}, where it is shown that the Wigner
functions go to KvN wavefunctions in a suitable limit. Other works
include the quantum-classical correspondence for integrable and
chaotic systems in the $h\rightarrow0$ limit \cite{Cla9} and geometric
dequantization \cite{ClaA}.  On the other hand, studies in the opposite direction can also be found, references \cite{cla 6a, cla 6} give procedures to obtain quantum mechanics from the operational classical mechanics. Moreover, the KvN theory
has also been used to derive purely classical results \cite{Cla 7,Cla 8,minimal coupling, Cla I, Cla P}.

The KvN theory starts with the Liouville equation, and from it a Hilbert
space and a set of relevant operators are built. We take a different
approach to get an operational formulation of classical mechanics.
Namely, in this article we derive an operational formulation from a unitary, irreducible representation of the Galilei
group. Unlike the KvN theory, our approach is independent of any previous
results from analytical mechanics. From the beginning we will postulate
a complex Hilbert space $\mathcal{H_{C}}$, and then we will look for a realization
of the Galilei group where the space-time transformations are represented
by unitary transformations acting on $\mathcal{H}_{C}$. We will show,
by direct construction, that this program is compatible with the basic
postulates of classical mechanics and that it contains the KvN theory
as a particular case. Our presentation will be close to the one given
in \cite{jordan,ballentine} for the derivation of quantum mechanics
from the Galilei algebra.

The organization of this article is as follows. In the following section
we give a brief summary of the Galilei group, and we present the conventions
that will be followed in the rest of this work. For the purpose of
this article we shall only need the most basic properties of the Galilei
group and algebra. For reasons of necessity, our notation for the
generators of the Galilei algebra is not the standard one used in
quantum mechanics.

In section 3 we introduce the Hilbert space $\mathcal{H}_{C}$ where
the operators from the Galilei group act upon, and we state the action
of these operators over the vectors of $\mathcal{H_{C}}$. There we
also define operators outside the generators of the Galilei group
that are necessary to give a physical interpretation of the theory,
namely the position and velocity operators. We will postulate the
basic relations between the operators of the theory that are consistent
with the requirements of a classical dynamic. We will then proceed
to find an irreducible unitary representation of the Galilei group
for the free particle and for a particle interacting with an external
force.

In section 4 we give an operational approach to concepts from analytical
mechanics like the Lagrangian and the canonical momentum. We will
show that the definition of a canonical momentum allows for an alternative,
but unitary equivalent, representation of the Galilei group. This
alternative representation is the Koopman-von Neumann theory. In section
5 we point out the relation between the operational approach to dynamics
with the Hamiltonian mechanics.

In section 6 we compare the irreducible unitary representations of
the Galilei group for classical and quantum mechanics.
We expect that the ab initio construction of a
Hilbert space associated with classical mechanics will help to reveal more
about the fundamental differences and similarities between the two theories.

The Einstein summation convention is used through all this work. 

\section{The Galilei Group and Algebra}

The proper Galilei group is a ten parameter group that consists of
space and time translations, rotations and pure Galilei transformation
(boosts). The general transformation $(\mathbf{x},t)\rightarrow(\mathbf{x}',t')$
can be written as 

\begin{eqnarray*}
\mathbf{x}' & = & R\mathbf{x}+\mathbf{b}t+\mathbf{a},\\
t' & = & t+\tau,
\end{eqnarray*}
where $R$ is a rotation, $\mathbf{a}$ is a space displacement, $\mathbf{b}$
is the velocity of a moving frame and $\tau$ is a time displacement.
The generators of the basic group transformations will be associated
with Hermitian operators as follows: $\widehat{\mathcal{J}}_{\alpha}$
stands for the rotations around $\alpha$-axis ($\alpha=1,2,3$);
$\widehat{\lambda}_{x_{\alpha}}$ is the space displacement generator
in the $\alpha$ -direction; $\widehat{\mathcal{G}}_{\alpha}$ correspond
to the Galilean boost along the $\alpha$-axis; $\widehat{L}$ will
be the time displacement generator. All these operators will act on
a suitable Hilbert space to be described in the next section. 

The space-time transformation of the Galilei group will be realized
by unitary operators with the following convention:

\begin{eqnarray*}
\mathrm{\mathbf{Space-Time\:Transformations}} &  & \mathbf{Unitary\:Operator}\\
\mathrm{Rotations} & \qquad & \mathrm{}\\
\mathbf{x}\rightarrow R_{\alpha}(\theta_{\alpha})\mathbf{x} & \qquad & e^{-i\theta_{\alpha}\widehat{\mathcal{J}}_{\alpha}}.\\
\mathrm{Spatial\:Displacement}\\
\mathbf{x}\rightarrow\mathbf{x}+\mathbf{a} &  & e^{-ia_{\alpha}\widehat{\lambda}_{x_{\alpha}}}.\\
\mathrm{Galilean\:Boost}\\
\mathbf{x}\rightarrow\mathbf{x}+\mathbf{b}t &  & e^{ib_{\alpha}\widehat{\mathcal{G}}_{\alpha}}.\\
\mathrm{Time\:Displacement}\\
t\longrightarrow t+\tau &  & e^{i\tau\widehat{L}}.
\end{eqnarray*}

The derivation of the Lie algebra associated to the Galilei group
can be found in many places, for example \cite{jordan}. It will be
useful to divide the Galilei algebra relations in two sets. The first
set does not involve the time displacement generator in the commutation
relations

\begin{subequations}

\begin{eqnarray}
\left[\widehat{\lambda}_{x_{\alpha}},\widehat{\lambda}_{x_{\beta}}\right] & = & 0,\\
\left[\widehat{\mathcal{G}}_{\alpha},\widehat{\mathcal{G}}_{\beta}\right] & = & 0,\\
\left[\widehat{\mathcal{J}}_{\alpha},\widehat{\mathcal{J}}_{\beta}\right] & = & i\varepsilon_{\alpha\beta\gamma}\widehat{\mathcal{J}}_{\gamma},\label{jj}\\
\left[\widehat{\mathcal{J}}_{\alpha},\widehat{\lambda}_{x_{\beta}}\right] & = & i\varepsilon_{\alpha\beta\gamma}\widehat{\lambda}_{x_{\gamma}},\label{jlambdax}\\
\left[\widehat{\mathcal{J}}_{\alpha},\widehat{\mathcal{G}}_{\beta}\right] & = & i\varepsilon_{\alpha\beta\gamma}\widehat{\mathcal{G}}_{\gamma},\\
\left[\widehat{\mathcal{G}}_{\alpha},\widehat{\lambda}_{x_{\beta}}\right] & = & i\delta_{\alpha\beta}\mathcal{M}.\label{Glambda}
\end{eqnarray}
\end{subequations}

The second is composed by the brackets that do involve $\widehat{L}$

\begin{subequations}

\begin{eqnarray}
\left[\widehat{\mathcal{J}}_{\alpha},\widehat{L}\right] & = & 0,\label{JL}\\
\left[\widehat{\mathcal{G}}_{\alpha},\widehat{L}\right] & = & i\widehat{\lambda}_{x_{\alpha}},\label{gl}\\
\left[\widehat{\lambda}_{x_{\alpha}},\widehat{L}\right] & = & 0.\label{lambdaxL}
\end{eqnarray}

\end{subequations}

In (\ref{Glambda}) $\mathcal{M}$ is the central charge of the algebra
and can be thought just as a real number whose value is not determined
a priori by any equation.

We are using the unusual symbols $(\widehat{\mathcal{J}},\,\widehat{\lambda}_{\mathbf{r}},\,\widehat{\mathcal{G}},\,\widehat{L},\,\mathcal{M})$
for the Galilei algebra, instead of the more common ones $(\widehat{\mathbf{J}},\,\widehat{\mathbf{P}},\,\widehat{\mathbf{G}},\,\widehat{H},\,M)$
used in quantum mechanics, because in our representation of the Galilei
group the elements of the Lie algebra, although related, will not
be identified with usual physical quantities. For example, the generator
of rotation $\widehat{\mathcal{J}}$ is not an angular momentum operator,
and $\widehat{L}$ is not an energy operator. Perhaps the most dramatic
difference is in the number $\mathcal{M}$ which will not be the mass
of the system, and, as we will show later, the situation for the allowed
value for $\mathcal{M}$ is completely different than in quantum mechanics
where the mass $M$ has to be a positive number in order to have a
physical representation of the Galilei group.

\section{Classical Representation of the Galilei Algebra}

We posit that the position $\mathbf{r}$ and the velocity $\mathbf{v}$
of a point particle can be simultaneously measured to any degree of
accuracy, and our formalism will reflect this. The state of the classical
particle will be described by a vector in a suitable complex Hilbert space.
To account for the lack of an uncertainty principle between the classical
particle's position and velocity (or later on, the momentum) the state
vectors should contain information of both the particle's position
and velocity\footnote{In this paper we ignore the possibility of internal degree of freedom
that come from the spin. }; hence, we postulate that the Hilbert space $\mathcal{H}_{C}$ of
the classical point particle is composed of vectors $\left|\psi\right\rangle $
of the form 

\begin{equation}
\left|\psi\right\rangle =\int \left\langle \mathbf{r},\mathbf{v}\right|\left.\psi\right\rangle \left|\mathbf{r},\mathbf{v}\right\rangle \, d\mathbf{r}d\mathbf{v},
\end{equation}
such that $\psi(\mathbf{r},\mathbf{v})=\left\langle \mathbf{r},\mathbf{v}\right.\left|\psi\right\rangle $
is a square integrable function and the kets $\left|\mathbf{r},\mathbf{v}\right\rangle $
obey the orthornormality condition

\begin{equation}
\left\langle \mathbf{r}',\mathbf{v}'\right.\left|\mathbf{r},\mathbf{v}\right\rangle =\delta(\mathbf{r}-\mathbf{r}')\delta(\mathbf{v}-\mathbf{v}').
\end{equation}
The probability $P(\mathbf{r},\mathbf{v})$ of finding the particle
with position $\mathbf{r}$ and velocity $\mathbf{v}$ is given by
the Born rule

\begin{equation}
P(\mathbf{r},\mathbf{v})=\left|\left\langle \mathbf{r},\mathbf{v}\right.\left|\Psi\right\rangle \right|^{2}.
\end{equation}
Geometrically, the effect of $\widehat{\lambda}_{\mathbf{r}}$, $\widehat{\mathcal{G}}$
and $\widehat{\mathbf{\mathcal{J}}}$ represent a translation in the
spatial coordinates, a Galilean boost and a rotation respectively.
Accordingly, we will demand the action of these operators on the base
kets to be

\begin{subequations}

\begin{eqnarray}
e^{-i\mathbf{a}\cdot\widehat{\lambda}_{\mathbf{r}}}\left|\mathbf{r},\mathbf{v}\right\rangle  & = & \left|\mathbf{r}+\mathbf{a},\mathbf{v}\right\rangle ,\label{Def lambdar}\\
e^{i\mathbf{b}\cdot\widehat{\mathcal{G}}}\left|\mathbf{r},\mathbf{v}\right\rangle  & \propto & \left|\mathbf{r}+\mathbf{b}t,\mathbf{v}+\mathbf{b}\right\rangle ,\label{Def G}\\
e^{-i\theta\hat{\mathbf{n}}\cdot\widehat{\mathbf{\mathcal{J}}}}\left|\mathbf{r},\mathbf{v}\right\rangle  & = & \left|\mathbf{r}+\theta\hat{\mathbf{n}}\times\mathbf{r},\mathbf{v}+\theta\hat{\mathbf{n}}\times\mathbf{v}\right\rangle ,\label{Def J}
\end{eqnarray}

\end{subequations}

where in (\ref{Def G}) we have used the proportionality sign to indicate
that a phase factor can be present due to commutation relation (\ref{Glambda}).
The effect of $\widehat{L}$ is temporal displacement in the state
vectors, namely

\begin{equation}
e^{-it\widehat{L}}\left|\Psi(0)\right\rangle =\left|\Psi(t)\right\rangle .\label{def time evolution}
\end{equation}

As in quantum mechanics, the time displacement equation (\ref{def time evolution})
implies a Schr\"odinger equation

\begin{eqnarray}
\frac{d}{dt}\left|\Psi(t)\right\rangle  & = & -i\widehat{L}\left|\Psi(t)\right\rangle .\label{defL2}
\end{eqnarray}
It is natural to introduce position and velocity operators $\widehat{\mathbf{R}}=\left(\widehat{X}_{1},\widehat{X}_{2},\widehat{X}_{3}\right)$
and $\widehat{\mathbf{V}}=\left(\widehat{V}_{1},\widehat{V}_{2},\widehat{V}_{3}\right)$
such that, by definition, we have

\begin{subequations}
\begin{eqnarray}
\widehat{X}_{\alpha}\left|\mathbf{r},\mathbf{v}\right\rangle  & = & x_{\alpha}\left|\mathbf{r},\mathbf{v}\right\rangle ,\\
\widehat{V}_{\alpha}\left|\mathbf{r},\mathbf{v}\right\rangle  & = & v_{\alpha}\left|\mathbf{r},\mathbf{v}\right\rangle .
\end{eqnarray}

\end{subequations}

We will also assume the existence of operators $\widehat{\lambda}_{\mathbf{v}}=\left(\widehat{\lambda}_{v_{1}},\widehat{\lambda}_{v_{2}},\widehat{\lambda}_{v_{3}}\right)$
that act as translation operators in the velocity coordinates

\begin{equation}
e^{-i\mathbf{b}\cdot\widehat{\lambda}_{\mathbf{v}}}\left|\mathbf{r},\mathbf{v}\right\rangle =\left|\mathbf{r},\mathbf{v}+\mathbf{b}\right\rangle .\label{DEf lambdav}
\end{equation}

As $\widehat{\lambda}_{x_{\alpha}}$ and $\widehat{\lambda}_{v_{\alpha}}$
are translation operators, they are conjugated in the quantum sense
to $\widehat{X}_{\alpha}$ and $\widehat{V}_{\alpha}$ respectively.
The following commutation relations are postulated to be satisfied

\begin{subequations}

\begin{eqnarray}
\left[\widehat{X}_{\alpha},\widehat{X}_{\beta}\right] & = & \left[\widehat{X}_{\alpha},\widehat{V}_{\beta}\right]=\left[\widehat{V}_{\alpha},\widehat{V}_{\beta}\right]=0,\label{XV}\\
\left[\widehat{X}_{\alpha},\widehat{\lambda}_{v_{\beta}}\right] & = & \left[\widehat{V}_{\alpha},\widehat{\lambda}_{x_{\beta}}\right]=0,\\
\left[\widehat{X}_{\alpha},\widehat{\lambda}_{x_{\beta}}\right] & = & i\delta_{\alpha\beta},\label{Xlambda}\\
\left[\widehat{V}_{\alpha},\widehat{\lambda}_{v_{\beta}}\right] & = & i\delta_{\alpha\beta}.\label{Vlambda}
\end{eqnarray}

\end{subequations}

Equations (\ref{XV}) summarize the physical statement of classical
mechanics that the position and the velocity of a particle can be
known with any desired degree of accuracy. 

The operators $\widehat{\mathbf{R}}$, $\widehat{\mathbf{V}}$, and
$\widehat{\lambda}_{\mathbf{v}}$ have to be vector operators; this
is, their components have to transform under rotations according to

\begin{subequations}

\begin{eqnarray}
\left[\widehat{\mathcal{J}}_{\alpha},\widehat{X}_{\beta}\right] & = & i\varepsilon_{\alpha\beta\gamma}\widehat{X}_{\gamma},\label{jx}\\
\left[\widehat{\mathcal{J}}_{\alpha},\widehat{V}_{\beta}\right] & = & i\varepsilon_{\alpha\beta\gamma}\widehat{V}_{\gamma},\label{jv}\\
\left[\widehat{\mathcal{J}}_{\alpha},\widehat{\lambda}_{v_{\beta}}\right] & = & i\varepsilon_{\alpha\beta\gamma}\widehat{\lambda}_{v_{\gamma}}.\label{jlambdav}
\end{eqnarray}

\end{subequations}

Furthermore, the effect of a Galilean boost on $\mathbf{\widehat{R}}$
and $\mathbf{\widehat{V}}$ is the same as in quantum mechanics \cite{ballentine},
i.e., the boost generates a displacement in the operators as follows

\begin{subequations}

\begin{eqnarray}
e^{i\mathbf{b}\cdot\widehat{\mathcal{G}}}\mathbf{\widehat{V}}e^{-i\mathbf{b}\cdot\widehat{\mathcal{G}}} & = & \mathbf{\widehat{V}}-\mathbf{b},\label{GV}\\
e^{i\mathbf{b}\cdot\widehat{\mathcal{G}}}\mathbf{\widehat{R}}e^{-i\mathbf{b}\cdot\widehat{\mathcal{G}}} & = & \mathbf{\widehat{R}}-\mathbf{b}t.\label{GR}
\end{eqnarray}

\end{subequations}

Let us note that (\ref{GV}) and (\ref{GR}) imply that

\begin{subequations}

\begin{eqnarray}
\left[\widehat{X}_{\alpha},\widehat{\mathcal{G}}_{\beta}\right] & = & i\delta_{\alpha\beta}t,\label{xcommg}\\
\left[\widehat{V}_{\alpha},\widehat{\mathcal{G}}_{\beta}\right] & = & i\delta_{\alpha\beta}.\label{vcommg}
\end{eqnarray}

\end{subequations}

We postulate the same dynamical relation between the position operator
$\widehat{X}_{\alpha}$ and $\widehat{V}_{\alpha}$ as in quantum
mechanics, namely \cite{ballentine}

\begin{equation}
\frac{d}{dt}\left\langle \widehat{\mathbf{R}}\right\rangle =\left\langle \widehat{\mathbf{V}}\right\rangle .\label{RV1}
\end{equation}

Due to equation (\ref{defL2}), equation (\ref{RV1}) reduces to

\begin{equation}
\widehat{\mathbf{V}}=i\left[\widehat{L},\widehat{\mathbf{R}}\right].\label{RV2}
\end{equation}

Finally, consider the acceleration operator $\widehat{\mathbf{a}}$
defined by

\begin{equation}
\frac{d}{dt}\left\langle \widehat{\mathbf{V}}\right\rangle =\left\langle \widehat{\mathbf{a}}\right\rangle .\label{VA1}
\end{equation}
We demand the acceleration operator to be function of $\widehat{\mathbf{R}}$
and $\widehat{\mathbf{V}}$, and to be independent of $\widehat{\lambda}_{\mathbf{v}}$
and $\widehat{\lambda}_{\mathbf{r}}$, i.e, $\widehat{\mathbf{a}}=\widehat{\mathbf{a}}(\widehat{\mathbf{R}},\widehat{\mathbf{V}})$.
In view of this last condition, equation (\ref{VA1}) imposes the
following on $\widehat{L}$

\begin{equation}
i\left[\widehat{L},\widehat{\mathbf{V}}\right]=\frac{1}{m}\widehat{\mathbf{F}}(\widehat{\mathbf{R}},\widehat{\mathbf{V}}),\label{LVF}
\end{equation}
for some function $\widehat{\mathbf{F}}$ where $m$ is to be identified
with the mass of the particle\footnote{The introduction here of a mass $m$ is rather artificial compared
to the quantum case where the mass appears in the Galilei algebra
and acts as a superselection operator. In classical mechanics the
concept of mass becomes important because, under the same force, it
is possible to observe different responses from different particles
. As we are dealing here with only one particle, it is of no surprise
that we could make $m$ disappear by a redefinition of the force as
the same can be done in Newtonian mechanics.}.

We shall see in section 5 the relation of $\widehat{L}$ with the
Liouville equation of classical statistical mechanics; for this reason,
we will call $\widehat{L}$ the Liouvillian of the system.

As we are ignoring internal degrees of freedom, the set of six operators
$\left\{ \widehat{\mathbf{R}},\widehat{\mathbf{V}}\right\} $ form
a complete set of commuting operators in the Hilbert space of a single
classical particle. Therefore, due to commutation relations (\ref{XV})
to (\ref{Vlambda}) and Schur's lemma \cite{jordan} the collection
$\left\{ \widehat{\mathbf{R}},\widehat{\mathbf{V}},\widehat{\lambda}_{\mathbf{r}},\widehat{\lambda}_{\mathbf{v}}\right\} $
forms an irreducible set of operators on the Hilbert space of the
square integrable wave functions $\psi(\mathbf{r},\mathbf{v})=\left\langle \mathbf{r},\mathbf{v}\right.\left|\psi\right\rangle $.
In the following section we are going to use the commutation relations
of the Galilei algebra together with the definitions presented in
this section to find a realization for $\widehat{\mathcal{G}}$, $\widehat{\mathbf{\mathcal{J}}}$
and $\widehat{L}$ in terms of $\left\{ \widehat{\mathbf{R}},\widehat{\mathbf{V}},\widehat{\lambda}_{\mathbf{r}},\widehat{\lambda}_{\mathbf{v}}\right\} $. 

\subsection{Free Particle}

The dynamic of the free particle has to be invariant under the full
Galilei group of space-time transformations. As it is done in quantum
mechanics, we will use this invariance to identify the generators
of the group.

We will find first the representation of the rotation operator $\widehat{\mathbf{\mathcal{J}}}$.
It can be checked by direct computation that the following operators

\begin{equation}
\widehat{\mathcal{J}}_{\alpha}=\varepsilon_{\alpha\beta\gamma}\left(\widehat{X}_{\beta}\widehat{\lambda}_{x_{\gamma}}+\widehat{V}_{\beta}\widehat{\lambda}_{v_{\gamma}}\right)\label{formJ}
\end{equation}
satisfy each of the conditions (\ref{jj}), (\ref{jlambdax}), (\ref{jx}),
(\ref{jv}) and (\ref{jlambdav}); signifying that we have a viable
representation for $\widehat{\mathbf{\mathcal{J}}}$. Now, let us
write $\widehat{\mathcal{G}}$ as

\begin{equation}
\widehat{\mathcal{G}}_{\alpha}=-\widehat{\lambda}_{x_{\alpha}}t-\widehat{\lambda}_{v_{\alpha}}+\mathbf{W},\label{Formg1}
\end{equation}
where $W_{\alpha}=W_{\alpha}(\widehat{\mathbf{R}},\widehat{\mathbf{V}})$
is an arbitrary function. Equation (\ref{Formg1}) satisfies each
of the conditions (\ref{Def G}), (\ref{GV}) and (\ref{GR}). Equation
(\ref{Glambda}) constrains $W_{\alpha}$ to be linear in $\widehat{X}_{\alpha}$.
The only vector function of $\widehat{\mathbf{V}}$ is itself, and
the only rotational invariant function of $\widehat{\mathbf{V}}$
is $\widehat{V}{}^{2}$. Thus, the possible $W_{\alpha}$ has to be
of the form 

\begin{equation}
W_{\alpha}=\mathcal{M}\widehat{X}_{\alpha}+w\left(\widehat{V}{}^{2}\right)\widehat{V}_{\alpha}.\label{GG0}
\end{equation}
We postpone further investigations of $\widehat{\mathcal{G}}_{\alpha}$
until we have examined the time displacement generator $\widehat{L}$.

Equation (\ref{RV2}) will be fulfilled if the Liouvillian is of the
form

\begin{equation}
\widehat{L}=\widehat{V}_{\alpha}\widehat{\lambda}_{x_{\alpha}}+f(\widehat{\mathbf{R}},\widehat{\mathbf{V}},\widehat{\lambda}_{\mathbf{v}}),
\end{equation}
where $f$ is, so far, an arbitrary function. We can use others commutators
to investigate the allowed functions $f$. For example, from (\ref{lambdaxL})
we can deduce that $f$ can not be a function of $\widehat{\mathbf{R}}$,
and from (\ref{LVF}) $f$ can be at most linear in $\widehat{\lambda}_{\mathbf{v}}$,
so we can write $\widehat{L}$ as

\begin{equation}
\widehat{L}=\widehat{V}_{\alpha}\widehat{\lambda}_{x_{\alpha}}+f_{\alpha}(\widehat{\mathbf{V}})\widehat{\lambda}_{v_{\alpha}}.\label{Lb}
\end{equation}
Now, $\widehat{\mathcal{G}}_{\alpha}$ and $\widehat{L}$ are related
by (\ref{gl}). Equations (\ref{Lb}) and (\ref{gl}) are incompatible
with $\widehat{\mathcal{G}}$ having a linear term in $\widehat{X}_{\alpha}$.
The only possibility to avoid a contradiction is to have $\mathcal{M}=0$
in equation (\ref{GG0}). Moreover, at least one of $f_{\alpha}$ and
$w_{\alpha}$ have to vanish in order to fulfill (\ref{gl}). We will
prove that $f_{\alpha}$ vanishes. We can write (\ref{gl}) as

\begin{equation}
[-\widehat{\lambda}_{v_{\alpha}},\widehat{L}]=i\widehat{\lambda}_{x_{\alpha}},
\end{equation}
or 

\begin{equation}
[\widehat{\lambda}_{v_{\alpha}},\widehat{L}-\widehat{V}_{\alpha}\widehat{\lambda}_{x_{\alpha}}]=0.\label{LNO}
\end{equation}
We can see in (\ref{LNO}) that $\widehat{L}-\widehat{V}_{\alpha}\widehat{\lambda}_{x_{\alpha}}$
is not a function of $\widehat{\mathbf{V}}$; therefore, $\widehat{L}$
have to be of the form

\begin{equation}
\widehat{L}=\widehat{V}_{\alpha}\widehat{\lambda}_{x_{\alpha}}+B_{\alpha}\widehat{\lambda}_{v_{\alpha}},
\end{equation}
where the $B_{\alpha}$ are constants. However, the term $B_{\alpha}\widehat{\lambda}_{v_{\alpha}}$
breaks the rotational invariance of $\widehat{L}$ given by equation
(\ref{JL}); hence, it can not be allowed. The final form of $\widehat{L}$
turns out to be

\begin{equation}
\widehat{L}=\widehat{V}_{\alpha}\widehat{\lambda}_{x_{\alpha}}.\label{freeL}
\end{equation}

It can be immediately checked that (\ref{freeL}) gives the expected
value of the acceleration for a free particle

\begin{equation}
\widehat{\mathbf{a}}=i\left[\widehat{L},\widehat{\mathbf{V}}\right]=0.\label{LVF-2}
\end{equation}

Under the action of $\widehat{L}$, the kets $\left|\mathbf{r},\mathbf{v}\right\rangle $
transform as expected for the free particle

\begin{eqnarray}
e^{-it\widehat{L}}\left|\mathbf{r},\mathbf{v}\right\rangle  & = & e^{-it\widehat{\mathbf{V}}\cdot\widehat{\lambda}_{\mathbf{r}}}\left|\mathbf{r},\mathbf{v}\right\rangle ,\nonumber \\
e^{-it\mathbf{v}\cdot\widehat{\lambda}_{\mathbf{r}}}\left|\mathbf{r},\mathbf{v}\right\rangle  & = & \left|\mathbf{r}+\mathbf{v}t,\mathbf{v}\right\rangle .
\end{eqnarray}

There is no commutation relation in the Lie Algebra of the Galilei
group, nor in the postulated relations for $\widehat{\mathbf{R}}$,
$\widehat{\mathbf{V}}$ and $\widehat{\lambda}_{\mathbf{v}}$ that
allow an unequivocally determination of the function $w$ in (\ref{GG0}).
The existence of $w$ would not affect any of the commutation relations
already defined, nor the dynamical equations that we will find later.
Since it seems that we have the freedom to choose any $w$ we want,
we will take the simplest choice $w=0$. The above imply the following
commutation relation

\begin{equation}
\left[\widehat{\mathcal{G}}_{\alpha},\widehat{\lambda}_{v_{\beta}}\right]=0.
\end{equation}

To recapitulate the results so far, as long as $\mathcal{M}=0$, the
elements of the Galilei algebra for the free particle can be written
in terms of the irreducible set of operators $\left\{ \widehat{\mathbf{R}},\widehat{\mathbf{V}},\widehat{\lambda}_{\mathbf{r}},\widehat{\lambda}_{\mathbf{v}}\right\} $
as

\begin{subequations}

\begin{eqnarray}
\widehat{\mathcal{J}}_{\alpha} & = & \varepsilon_{\alpha\beta\gamma}\left(\widehat{X}_{\beta}\widehat{\lambda}_{x_{\gamma}}+\widehat{V}_{\beta}\widehat{\lambda}_{v_{\gamma}}\right),\\
\widehat{\mathcal{G}}_{\alpha} & = & -\widehat{\lambda}_{x_{\alpha}}t-\widehat{\lambda}_{v_{\alpha}},\label{GG}\\
\widehat{L} & = & \widehat{V}_{\alpha}\widehat{\lambda}_{x_{\alpha}}.
\end{eqnarray}

\end{subequations}

\subsection{A Particle Interacting With External Forces}

An interaction with an external force modifies the time evolution
of the state vector compared to the free particle case. As in quantum
mechanics, we retain the equation of motion

\begin{eqnarray}
\frac{d}{dt}\left|\Psi(t)\right\rangle  & = & -i\widehat{L}\left|\Psi(t)\right\rangle, \label{defL2-1}
\end{eqnarray}
but we modify $\widehat{L}$ so that it will accounts for the interactions.
The operator $\widehat{L}$ is now to be understood as the operator
of \emph{dynamical} evolution in time. As $\widehat{L}$ is changed,
the commutation relationships (\ref{JL}), (\ref{gl}) and (\ref{lambdaxL})
can no longer be used. 

The transformations generated by $\widehat{\mathcal{G}}$ and $\widehat{\mathbf{\mathcal{J}}}$
are understood to be purely geometrical in nature, so the expressions
for them found in the previous section remain the same.

To identify the form of $\widehat{L}$, we will use the dynamical
equations

\begin{eqnarray}
i\left[\widehat{L},\widehat{X}_{\alpha}\right] & = & \widehat{V}_{\alpha},\label{LXV}\\
i\left[\widehat{L},\widehat{V}_{\alpha}\right] & = & \frac{1}{m}\widehat{F}_{\alpha}(\widehat{\mathbf{R}},\widehat{\mathbf{V}}).\label{LVF-1}
\end{eqnarray}

It can be readily checked that

\begin{equation}
\widehat{L}=\widehat{V}_{\alpha}\widehat{\lambda}_{x_{\alpha}}+\frac{1}{2m}\left(\widehat{F}_{\alpha}\widehat{\lambda}_{v_{\alpha}}+\widehat{\lambda}_{v_{\alpha}}\widehat{F}_{\alpha}\right)\label{forceL}
\end{equation}
fulfills both equations (\ref{LXV}) and (\ref{LVF-1}). Using equation
(\ref{forceL}), the commutators of $\widehat{L}$ with $\widehat{\lambda}_{\mathbf{r}}$
and $\widehat{\lambda}_{\mathbf{v}}$ can be computed as

\begin{eqnarray}
i\left[\widehat{L},\widehat{\lambda}_{x_{\alpha}}\right] & = & -\frac{1}{2m}\left(\frac{\partial\widehat{F}_{\beta}}{\partial\widehat{X}_{\alpha}}\widehat{\lambda}_{v_{\beta}}+\widehat{\lambda}_{v_{\beta}}\frac{\partial\widehat{F}_{\beta}}{\partial\widehat{X}_{\alpha}}\right),\\
i\left[\widehat{L},\widehat{\lambda}_{v_{\alpha}}\right] & = & -\widehat{\lambda}_{x_{\alpha}}-\frac{1}{2m}\left(\frac{\partial\widehat{F}_{\beta}}{\partial\widehat{V}_{\alpha}}\widehat{\lambda}_{v_{\beta}}+\widehat{\lambda}_{v_{\beta}}\frac{\partial\widehat{F}_{\beta}}{\partial\widehat{V}_{\alpha}}\right).
\end{eqnarray}

As we already have expressed $\widehat{\mathcal{G}}$ and $\widehat{\mathbf{\mathcal{J}}}$
in terms of the irreducible set $\left\{ \widehat{\mathbf{R}},\widehat{\mathbf{V}},\widehat{\lambda}_{\mathbf{r}},\widehat{\lambda}_{\mathbf{v}}\right\} $,
their commutators with $\widehat{L}$ can be computed, and the result
will depend on the given $\widehat{\mathbf{F}}$.

The form of the force operator is not given a priori, $\widehat{\mathbf{F}}$
has to be investigated from the experiments. Once $\widehat{\mathbf{F}}$
is found, the evolution in the state of the particle can be known
at any time solving the Schr\"odinger-like equation (\ref{defL2-1}).

\subsection*{Example: Projectile Motion}

As an example of the formalism developed above, let us consider the
case of projectile motion. For this case the force operator is

\begin{equation}
\widehat{\mathbf{F}}=-mg\hat{\mathbf{j}},
\end{equation}

where $g$ is a constant.

Using the formula (\ref{forceL}), the Liouvillian of this simple
system takes the form

\begin{equation}
\widehat{L}=\widehat{V}_{x}\widehat{\lambda}_{x}+\widehat{V}_{y}\widehat{\lambda}_{y}+\widehat{V}_{z}\widehat{\lambda}_{z}-g\widehat{\lambda}_{v_{y}}.
\end{equation}

As the Liouvillian is time independent, we can write the evolution
operator as follows

\begin{eqnarray}
U(t) & = & e^{-it\widehat{L}}=e^{-it\left(\widehat{V}_{x}\widehat{\lambda}_{x}+\widehat{V}_{y}\widehat{\lambda}_{y}+\widehat{V}_{z}\widehat{\lambda}_{z}-g\widehat{\lambda}_{v_{y}}\right)}\nonumber \\
 & = & e^{-it\left(\widehat{V}_{x}\widehat{\lambda}_{x}+\widehat{V}_{z}\widehat{\lambda}_{z}\right)}e^{-it\widehat{V}_{y}\widehat{\lambda}_{y}}e^{itg\widehat{\lambda}_{v_{y}}}e^{+i\frac{t^{2}}{2}\widehat{\lambda}_{y}},
\end{eqnarray}

where in the last step we used the well known operator formula $e^{t(A+B)}=e^{tA}e^{tB}e^{\frac{t^{2}}{2}\left[A,B\right]}$
valid when $A$ and $B$ commute with $\left[A,B\right]$ \cite{cohen}.

The effect of the evolution operator on a given base ket is

\begin{equation}
\left|\mathbf{r}(t);\mathbf{v}(t)\right\rangle =U(t)\left|\mathbf{r}_{0};\mathbf{v}_{0}\right\rangle =\bigl|\mathbf{r}_{0}+\mathbf{v}_{0}t-\frac{t^{2}}{2}g\hat{\mathbf{j}};\,\mathbf{v}_{0}-gt\hat{\mathbf{j}}\bigr\rangle,
\end{equation}

which is the expected behavior from a projectile motion. An arbitrary
wave function $\psi(\mathbf{r},\mathbf{v})=\bigl\langle\mathbf{r},\mathbf{v}\bigr|\psi\bigr\rangle$
will evolve in time according to the rule

\begin{eqnarray}
\psi(\mathbf{r},\mathbf{v},t) & = & \bigl\langle\mathbf{r}-\mathbf{v}t+\frac{t^{2}}{2}g\hat{\mathbf{j}};\,\mathbf{v}+gt\hat{\mathbf{j}}\bigr|\psi\bigr\rangle\nonumber \\
 & = & \psi(\mathbf{r}-\mathbf{v}t+\frac{t^{2}}{2}g\hat{\mathbf{j}};\,\mathbf{v}+gt\hat{\mathbf{j}}).
\end{eqnarray}

\subsection*{Example 2: Harmonic Oscillator}

For the sake of simplicity, we only consider the movement along $x$-axis
of a particle under the effect of a linear restoring force. Hence,
the base kets to be considered are of the form $\left|x,v_{x}\right\rangle $
and the force operator is given by

\begin{equation}
\widehat{F}=-k\widehat{X}.\label{FX}
\end{equation}

For the force (\ref{FX}) the Liouvillian takes the form

\begin{equation}
\widehat{L}=\widehat{V}_{x}\widehat{\lambda}_{x}-\omega^{2}\widehat{X}\widehat{\lambda}_{v_{x}},
\end{equation}
where $\omega^{2}=k/m$.

By means of the following transformation

\begin{eqnarray}
\widehat{V}_{x}' & = & \frac{1}{\omega}\widehat{V}_{x},\\
\widehat{\lambda}_{v_{x}}' & = & \omega\widehat{\lambda}_{v_{x}},
\end{eqnarray}
we can write the evolution operator as

\begin{equation}
U(t)=\exp\left[-i\omega t\left(\widehat{V}_{x}'\widehat{\lambda}_{x}-\widehat{X}\widehat{\lambda}_{v_{x}}'\right)\right].
\end{equation}

Algebraically, the combination $\widehat{V}_{x}'\widehat{\lambda}_{x}-\widehat{X}\widehat{\lambda}_{v_{x}}'$
have the same commutation relations with $\widehat{X}$ and $\widehat{V}_{x}$
as the $z$ component of the orbital angular momentum $J_{z}=\left(\widehat{\mathbf{r}}\times\widehat{\mathbf{p}}\right)_{z}$
has with the position operators $\widehat{x}$ and $\widehat{y}$
in quantum mechanics. Therefore, $\widehat{V}_{x}'\widehat{\lambda}_{x}-\widehat{X}\widehat{\lambda}_{v_{x}}'$
acts as a generator of rotations in a plane where the velocity plays
the role of a perpendicular axis to the $x$-axis., i.e., the tangent
bundle of configuration space, given in this case by the Cartesian
product $X\times V$. Hence, we can write the evolution of a base
ket as follows

\begin{equation}
U(t)\left|x_{0};v_{0}\right\rangle =\left|x_{0}\cos\left(\omega t\right)+\frac{v_{0}}{\omega}\sin\left(\omega t\right);-\omega x_{0}\sin\left(\omega t\right)+v_{0}\cos\left(\omega t\right)\right\rangle .
\end{equation}

\subsection{System of Mutually Interacting Particles.}

We obtained the operators that represent the classical dynamical variables
of a single particle, and now we are going to generalize the result
for the case of a system of interacting particles. We work out in
detail the case of two interacting particles. The case of N particles
will be just a trivial generalization.

Let the Hilbert spaces of particle 1 and particle 2 be $\mathcal{H}_{1}$
and $\mathcal{H}_{2}$, respectively. The set $\left\{ \widehat{\mathbf{R}}^{(1)},\widehat{\mathbf{V}}^{(1)},\widehat{\lambda}_{\mathbf{r}}^{(1)},\widehat{\lambda}_{\mathbf{v}}^{(1)}\right\} $
acts on $\mathcal{H}_{1}$ while $\left\{ \widehat{\mathbf{R}}^{(2)},\widehat{\mathbf{V}}^{(2)},\widehat{\lambda}_{\mathbf{r}}^{(2)},\widehat{\lambda}_{\mathbf{v}}^{(2)}\right\} $
acts on $\mathcal{H}_{2}$. Just as in quantum mechanics, and for
the same reasons \cite{ballentine}, we postulate that the Hilbert
space of the composite system is given by the tensor product

\begin{equation}
\mathcal{H}_{12}=\mathcal{H}_{1}\otimes\mathcal{H}_{2}.
\end{equation}

The set 

\begin{equation}
\left\{ \widehat{\mathbf{R}}^{(1)},\widehat{\mathbf{V}}^{(1)},\widehat{\mathbf{R}}^{(2)},\widehat{\mathbf{V}}^{(2)},\widehat{\lambda}_{\mathbf{r}}^{(1)},\widehat{\lambda}_{\mathbf{v}}^{(1)},\widehat{\lambda}_{\mathbf{r}}^{(2)},\widehat{\lambda}_{\mathbf{v}}^{(2)}\right\} 
\end{equation}
is irreducible on $\mathcal{H}_{12}$. The Liouvillian of the composite
system is postulated to take the form

\begin{equation}
\widehat{L}=\widehat{V}_{\alpha}^{(1)}\widehat{\lambda}_{x_{\alpha}}^{(1)}+\widehat{V}_{\alpha}^{(2)}\widehat{\lambda}_{x_{\alpha}}^{(2)}+W,
\end{equation}

where $W$ is an interaction term. Now, we demand that the interaction
term has to give rise to the force equation for the individual particles,
more precisely

\begin{eqnarray}
i\left[\widehat{L},\widehat{\mathbf{V}}^{(1)}\right] & = & \frac{1}{m_{1}}\widehat{\mathbf{F}}^{(1)},\label{F1}\\
i\left[\widehat{L},\widehat{\mathbf{V}}^{(2)}\right] & = & \frac{1}{m_{2}}\widehat{\mathbf{F}}^{(2)},\label{F2}
\end{eqnarray}
with $\widehat{\mathbf{F}}^{(1)}$ and $\widehat{\mathbf{F}}^{(2)}$
function of $\left\{ \widehat{\mathbf{R}}^{(1)},\widehat{\mathbf{V}}^{(1)},\widehat{\mathbf{R}}^{(2)},\widehat{\mathbf{V}}^{(2)}\right\} $
only. It can be checked that (\ref{F1}) and (\ref{F2}) are satisfied
by

\begin{eqnarray}
\widehat{L} & = & \widehat{V}_{\alpha}^{(1)}\widehat{\lambda}_{x_{\alpha}}^{(1)}+\widehat{V}_{\alpha}^{(2)}\widehat{\lambda}_{x_{\alpha}}^{(2)}+\frac{1}{2m_{1}}\left(\widehat{F}_{\alpha}^{(1)}\widehat{\lambda}_{v_{\alpha}}^{(1)}+\widehat{\lambda}_{v_{\alpha}}^{(1)}\widehat{F}_{\alpha}^{(1)}\right)\nonumber \\
 &  & +\frac{1}{2m_{2}}\left(\widehat{F}_{\alpha}^{(2)}\widehat{\lambda}_{v_{\alpha}}^{(2)}+\widehat{\lambda}_{v_{\alpha}}^{(2)}\widehat{F}_{\alpha}^{(2)}\right).
\end{eqnarray}
We want to know what are the constraints imposed on $\widehat{\mathbf{F}}^{(1)}$
and $\widehat{\mathbf{F}}^{(2)}$ by Galilean invariance. As in quantum
mechanics, the other elements of the Galilei algebra are the sum of
the individual generators for each particle, namely

\begin{eqnarray}
\widehat{\lambda}_{\mathbf{r}} & = & \widehat{\lambda}_{\mathbf{r}}^{(1)}+\widehat{\lambda}_{\mathbf{r}}^{(2)},\\
\widehat{\mathcal{J}}_{\alpha} & = & \widehat{\mathcal{J}}_{\alpha}^{(1)}+\widehat{\mathcal{J}}_{\alpha}^{(2)},\\
\widehat{\mathcal{G}}_{\alpha} & = & \widehat{\mathcal{G}}_{\alpha}^{(1)}+\widehat{\mathcal{G}}_{\alpha}^{(2)}.\label{Gcomp}
\end{eqnarray}

Although it does not belong to the Galilei algebra, we also define
an operator for the total translation in the velocities given by

\begin{equation}
\widehat{\lambda}_{\mathbf{v}}=\widehat{\lambda}_{\mathbf{v}}^{(1)}+\widehat{\lambda}_{\mathbf{v}}^{(2)}.
\end{equation}

Instead of the operators for the individual particles, let us consider
the center of mass and the velocity of the center of mass operators
given by (where $M=m_{1}+m_{2}$)

\begin{eqnarray}
\widehat{\mathbf{R}}_{CM} & = & \frac{1}{M}\left(m_{1}\widehat{\mathbf{R}}^{(1)}+m_{2}\widehat{\mathbf{R}}^{(2)}\right),\\
\widehat{\mathbf{V}}_{CM} & = & \frac{1}{M}\left(m_{1}\widehat{\mathbf{V}}^{(1)}+m_{2}\widehat{\mathbf{V}}^{(2)}\right),
\end{eqnarray}
the relative position and relative velocity operators according to

\begin{eqnarray}
\widehat{\mathbf{R}}_{Rel} & = & \widehat{\mathbf{R}}^{(1)}-\widehat{\mathbf{R}}^{(2)},\\
\widehat{\mathbf{V}}_{Rel} & = & \widehat{\mathbf{V}}^{(1)}-\widehat{\mathbf{V}}^{(2)},
\end{eqnarray}
and the relative translation operators

\begin{eqnarray}
\widehat{\lambda}_{\mathbf{r}}^{\star} & = & \frac{1}{M}\left(m_{2}\widehat{\lambda}_{\mathbf{r}}^{(1)}-m_{1}\widehat{\lambda}_{\mathbf{r}}^{(2)}\right),\\
\widehat{\lambda}_{\mathbf{v}}^{\star} & = & \frac{1}{M}\left(m_{2}\widehat{\lambda}_{\mathbf{v}}^{(1)}-m_{1}\widehat{\lambda}_{\mathbf{v}}^{(2)}\right).
\end{eqnarray}

It can be easily checked that the only non-vanishing commutators between
the elements of the set $\left\{ \widehat{\mathbf{R}}_{CM},\widehat{\mathbf{V}}_{CM},\widehat{\mathbf{R}}_{Rel},\widehat{\mathbf{V}}_{Rel},\widehat{\lambda}_{\mathbf{r}},\widehat{\lambda}_{\mathbf{v}},\widehat{\lambda}_{\mathbf{r}}^{\star},\widehat{\lambda}_{\mathbf{v}}^{\star}\right\} $
are

\begin{eqnarray}
\left[\widehat{X}_{CM,\alpha},\widehat{\lambda}_{x_{\alpha}}\right] & = & \left[\widehat{V}_{CM,\alpha},\widehat{\lambda}_{v_{\alpha}}\right]=\left[\widehat{X}_{Rel,\alpha},\widehat{\lambda}_{x_{\alpha}}^{\star}\right]\nonumber \\
 & = & \left[\widehat{V}_{Rel,\alpha},\widehat{\lambda}_{v_{\alpha}}^{\star}\right]=i.\label{rel}
\end{eqnarray}

We will use the set $\left\{ \widehat{\mathbf{R}}_{CM},\widehat{\mathbf{V}}_{CM},\widehat{\mathbf{R}}_{Rel},\widehat{\mathbf{V}}_{Rel},\widehat{\lambda}_{\mathbf{r}},\widehat{\lambda}_{\mathbf{v}},\widehat{\lambda}_{\mathbf{r}}^{\star},\widehat{\lambda}_{\mathbf{v}}^{\star}\right\} $
to study the allowed forms of $\widehat{\mathbf{F}}^{(1)}$ and $\widehat{\mathbf{F}}^{(2)}$
.

In view of the relations (\ref{JL}),(\ref{gl}) and (\ref{lambdaxL}),
the interaction term $W$  has to obey the following commutation relations

\begin{subequations}

\begin{eqnarray}
\left[\widehat{\lambda}_{\mathbf{r}},W\right] & = & 0,\label{w1}\\
\left[\widehat{\mathcal{G}}_{\alpha},W\right] & = & 0,\label{w2}\\
\left[\widehat{\mathcal{J}}_{\alpha},W\right] & = & 0.\label{w3}
\end{eqnarray}

\end{subequations}

Furthermore, the forces $\widehat{\mathbf{F}}^{(1)}$ and $\widehat{\mathbf{F}}^{(2)}$
should satisfy the commutation relations

\begin{eqnarray}
\left[\widehat{\lambda}_{\mathbf{r}},\widehat{\mathbf{F}}^{(1)}\right]=\left[\widehat{\lambda}_{\mathbf{r}},\widehat{\mathbf{F}}^{(2)}\right] & = & 0,\label{w1-1}\\
\left[\widehat{\mathcal{G}}_{\alpha},\widehat{\mathbf{F}}^{(1)}\right]=\left[\widehat{\mathcal{G}}_{\alpha},\widehat{\mathbf{F}}^{(2)}\right] & = & 0.\label{w2-1}
\end{eqnarray}

Equation (\ref{w1}) implies that the forces cannot be function of
$\widehat{\mathbf{R}}_{CM}$, as $\widehat{\mathbf{R}}_{CM}$ does
not commute with $\widehat{\lambda}_{\mathbf{r}}$. Using (\ref{Gcomp}),
equation (\ref{w2}) implies that the forces cannot be function of
$\widehat{\mathbf{V}}_{CM}$. Therefore, $\widehat{\mathbf{F}}^{(1)}$
and $\widehat{\mathbf{F}}^{(2)}$ can only be function of $\widehat{\mathbf{R}}_{Rel}$
and $\widehat{\mathbf{V}}_{Rel}$. Finally, the condition (\ref{w3})
is met only if the forces are vector operators.

The generalization for the case of N interacting particles is straightforward.
The Hilbert space of the composite system is the tensor product of
the individual Hilbert spaces. The interaction forces are limited
to be functions of the scalar combination of the relative positions
$\widehat{\mathbf{R}}^{(i)}-\widehat{\mathbf{R}}^{(j)}$ and the relative
velocities $\widehat{\mathbf{V}}^{(i)}-\widehat{\mathbf{V}}^{(j)}$.

\section{Lagrangian Operator}

In the last section we completed our task of giving a complete formulation
of the classical dynamics of a point particle. The developments of
this section are aimed to justify the definition of a canonical momentum
operator, as this new operator will help us to give an alternative
operational formulation of the classical dynamics. 

Let us start by noting that commutator (\ref{Vlambda}) allows us
to rewrite equation (\ref{LVF-1}) as

\begin{equation}
-\left[\widehat{L},\left[\widehat{\lambda}_{v_{\alpha}},\frac{m}{2}\widehat{V}^{2}\right]\right]=\widehat{F}_{\alpha}.\label{LVF-3}
\end{equation}
Moreover, let us decompose the forces as follows

\begin{equation}
\widehat{\mathbf{F}}=\widehat{\mathbf{F}}^{(C)}+\widehat{\mathbf{F}}^{(NC)},
\end{equation}
where, by definition, the components of the conservative forces can
be computed from a potential $U=U(\widehat{\mathbf{R}},\widehat{\mathbf{V}})$
according to

\begin{equation}
\widehat{F}_{\alpha}^{(C)}=-i\left[\widehat{\lambda}_{x_{\alpha}},\widehat{U}\right]-\left[\widehat{L},\left[\widehat{\lambda}_{v_{\alpha}},\widehat{U}\right]\right].\label{fpotential}
\end{equation}

With the help of (\ref{fpotential}), equation (\ref{LVF-3}) can
be further rewritten as

\begin{equation}
\Phi[\widehat{\mathcal{L}}]=\widehat{F}_{\alpha}^{(NC)},\label{phiL}
\end{equation}
where $\widehat{\mathcal{L}}$ is the Lagrangian operator

\begin{equation}
\widehat{\mathcal{L}}=\frac{m}{2}\widehat{V}^{2}-\widehat{U},\label{lagrangian}
\end{equation}
and $\Phi$ is the superoperator given by

\begin{equation}
\Phi=-\left[\widehat{L},\left[\widehat{\lambda}_{v_{\alpha}},\bullet\right]\right]-i\left[\widehat{\lambda}_{x_{\alpha}},\bullet\right],\label{superoperator}
\end{equation}
where the superoperator acts on $\widehat{\mathcal{L}}$ as $\Phi[\widehat{\mathcal{L}}]=-\left[\widehat{L},\left[\widehat{\lambda}_{v_{\alpha}},\widehat{\mathcal{L}}\right]\right]-i\left[\widehat{\lambda}_{x_{\alpha}},\widehat{\mathcal{L}}\right]$.

Equation (\ref{phiL}) is the equivalent of the Lagrange equations
in analytical mechanics.
From (\ref{fpotential}), the force $\widehat{\mathbf{F}}^{(C)}$
will be independent of the acceleration $\widehat{a}_{\alpha}=\left[\widehat{L},\widehat{V}_{\alpha}\right]$
only if $\widehat{U}$ is at most linear in the velocities. The generalized
potential is then of the form

\begin{equation}
\widehat{U}=\widehat{\phi}-\widehat{V}_{\alpha}\widehat{A}_{\alpha},\label{formofU}
\end{equation}
where both the scalar and vector potentials $\widehat{\phi}$ and
$\widehat{A}_{\alpha}$ are functions of $\widehat{\mathbf{R}}$ only

\begin{eqnarray}
\widehat{\phi} & = & \widehat{\phi}\left(\widehat{\mathbf{R}}\right),\\
\mathbf{\widehat{A}} & = & \mathbf{\widehat{A}}\left(\widehat{\mathbf{R}}\right).
\end{eqnarray}
Inserting the generalized potential (\ref{formofU}) into equation
(\ref{fpotential}) yields

\begin{eqnarray}
\widehat{F}_{\alpha}^{(C)} & = & -i\left[\widehat{\lambda}_{x_{\alpha}},U\right]-\left[\widehat{L},\left[\widehat{\lambda}_{v_{\alpha}},U\right]\right]+i\frac{\partial}{\partial t}\left[\widehat{\lambda}_{v_{\alpha}},U\right]\nonumber \\
 & = & -\frac{\partial\widehat{\phi}}{\partial\widehat{X}_{\alpha}}+i\left[\widehat{L},\widehat{A}_{\alpha}\right]\nonumber \\
 & = & -\frac{\partial\widehat{\phi}}{\partial\widehat{X}_{\alpha}}-\frac{\partial\widehat{A}_{\alpha}}{\partial t}-\widehat{V}_{\beta}\frac{\partial\widehat{A}_{\beta}}{\partial\widehat{X}_{\alpha}}+\widehat{V}_{\beta}\frac{\partial\widehat{A}_{\alpha}}{\partial\widehat{X}_{\beta}}\nonumber \\
 & = & \widehat{E}_{\alpha}+\left(\widehat{\mathbf{V}}\times\widehat{\mathbf{B}}\right)_{\alpha},
\end{eqnarray}

where

\begin{eqnarray}
\widehat{E}_{\alpha} & = & -\frac{\partial\widehat{\phi}}{\partial\widehat{X}_{\alpha}}-\frac{\partial\widehat{A}_{\alpha}}{\partial t},\\
\widehat{B}_{\alpha} & = & \left(\nabla\times\widehat{\mathbf{A}}\right)_{\alpha}.
\end{eqnarray}

There is a result in classical mechanics \cite{Santilli},  rediscovered by Feynman \cite{Fey} in a different context,  that states that
all the velocity dependent but acceleration independent forces that
can be derived from a Lagrangian have the form of the Lorentz force.  We have obtained here the same result starting from  
 different premises. 

\subsection{Momentum Representation}

Having the Lagrangian operator (\ref{lagrangian}), it is natural
to look for a definition of a canonical momentum operator $\widehat{\mathbf{P}}$.
We define the components of the canonical momentum operator by

\begin{equation}
\widehat{P}_{\alpha}=i\left[\widehat{\lambda}_{v_{\alpha}},\widehat{\mathcal{L}}\right]=m\widehat{V}_{\alpha}+\widehat{A}_{\alpha}.\label{defP}
\end{equation}
It is straightforward to check the following commutation relations
for $\widehat{\mathbf{P}}$

\begin{subequations}

\begin{eqnarray}
\left[\widehat{X}_{\alpha},\widehat{P}_{\beta}\right] & = & 0,\\
\left[\widehat{P}_{\alpha},\widehat{\lambda}_{v_{\beta}}\right] & = & im\delta_{\alpha\beta},\\
\left[\widehat{P}_{\alpha},\widehat{\lambda}_{x_{\beta}}\right] & = & i\frac{\partial\widehat{A}_{\alpha}}{\partial\widehat{X}_{\beta}}.
\end{eqnarray}

\end{subequations}

The introduction of a canonical momentum allows for a change of irreducible
representation of the Galilei group via the definitions

\begin{subequations}

\begin{eqnarray}
\widehat{\lambda}_{p_{\alpha}} & = & \frac{1}{m}\widehat{\lambda}_{v_{\alpha}},\label{changelambda1}\\
\widehat{\lambda}'_{x_{\beta}} & = & \widehat{\lambda}_{x_{\beta}}-\frac{\partial\widehat{A}_{\alpha}}{\partial\widehat{X}_{\beta}}\widehat{\lambda}_{p_{\alpha}}.\label{changelambda2}
\end{eqnarray}

\end{subequations}

The commutation relations between $\widehat{\mathbf{R}}$, $\widehat{\mathbf{P}}$,
$\widehat{\lambda}'_{\mathbf{r}}$ and $\widehat{\lambda}_{\mathbf{p}}$
are

\begin{subequations}

\begin{eqnarray}
\left[\widehat{X}_{\alpha},\widehat{X}_{\beta}\right] & = & \left[\widehat{X}_{\alpha},\widehat{P}_{\beta}\right]=\left[\widehat{P}_{\alpha},\widehat{P}_{\beta}\right]=0,\label{Xp}\\
\left[\widehat{X}_{\alpha},\widehat{\lambda}_{p\beta}\right] & = & \left[\widehat{P}_{\alpha},\widehat{\lambda}'_{x_{\beta}}\right]=0,\\
\left[\widehat{X}_{\alpha},\widehat{\lambda}'_{x_{\beta}}\right] & = & i\delta_{\alpha\beta},\label{Xlambda-1}\\
\left[\widehat{P}_{\alpha},\widehat{\lambda}_{p_{\beta}}\right] & = & i\delta_{\alpha\beta}.\label{Xp2}
\end{eqnarray}

\end{subequations}

The set of operators $\left\{ \widehat{\mathbf{R}},\widehat{\mathbf{P}},\widehat{\lambda}'_{\mathbf{r}},\widehat{\lambda}_{\mathbf{p}}\right\} $
is irreducible in the Hilbert space we are considering in view of
the preceding set of equations (\ref{Xp}) to (\ref{Xp2}). Therefore,
the transformations given by (\ref{defP}) , (\ref{changelambda1})
and (\ref{changelambda2}) allow us to pass from an irreducible representation
of the Galilei group in terms of $\left\{ \widehat{\mathbf{R}},\widehat{\mathbf{V}},\widehat{\lambda}_{\mathbf{r}},\widehat{\lambda}_{\mathbf{v}}\right\} $
to one given in terms of the operators $\left\{ \widehat{\mathbf{R}},\widehat{\mathbf{P}},\widehat{\lambda}'_{\mathbf{r}},\widehat{\lambda}_{\mathbf{p}}\right\} $.

The change from $\left\{ \widehat{\mathbf{R}},\widehat{\mathbf{V}},\widehat{\lambda}_{\mathbf{r}},\widehat{\lambda}_{\mathbf{v}}\right\} $
to $\left\{ \widehat{\mathbf{R}},\widehat{\mathbf{P}},\widehat{\lambda}'_{\mathbf{r}},\widehat{\lambda}_{\mathbf{p}}\right\} $
is a quantum canonical transformation as discussed in \cite{moshinsky}. This transformation can be done in two steps. First, we make the change

\begin{subequations}

\begin{eqnarray}
\widehat{\mathbf{V}} & \longrightarrow & m\widehat{\mathbf{V}},\\
\widehat{\lambda}_{\mathbf{v}} & \longrightarrow & \frac{1}{m}\widehat{\lambda}_{\mathbf{v}}.
\end{eqnarray}

\end{subequations}

Second, we will show that the set $\left\{ \widehat{\mathbf{R}},m\widehat{\mathbf{V}},\widehat{\lambda}_{\mathbf{r}},\frac{1}{m}\widehat{\lambda}_{\mathbf{v}}\right\} $
can be transformed into $\left\{ \widehat{\mathbf{R}},\widehat{\mathbf{P}},\widehat{\lambda}'_{\mathbf{r}},\widehat{\lambda}_{\mathbf{p}}\right\} $
by an unitary transformation using the unitary operator $\widehat{C}$
given by

\begin{equation}
\widehat{C}=e^{\frac{i}{m}\mathbf{\widehat{A}}\cdot\widehat{\lambda}_{\mathbf{v}}}.\label{CC}
\end{equation}

The above assertion can be checked by direct computation. The unitary
transformation given by the operator (\ref{CC}) leaves $\widehat{\mathbf{R}}$
and $\frac{1}{m}\widehat{\lambda}_{\mathbf{v}}$ unchanged,

\begin{subequations}

\begin{eqnarray}
e^{\frac{i}{m}\mathbf{\widehat{A}}\cdot\widehat{\lambda}_{\mathbf{v}}}\widehat{X}_{\alpha}e^{-\frac{i}{m}\mathbf{\widehat{A}}\cdot\widehat{\lambda}_{\mathbf{v}}} & = & \widehat{X}_{\alpha},\\
e^{-\frac{i}{m}\mathbf{\widehat{A}}\cdot\widehat{\lambda}_{\mathbf{v}}}\left(\frac{1}{m}\widehat{\lambda}_{v_{\alpha}}\right)e^{-\frac{i}{m}\mathbf{\widehat{A}}\cdot\widehat{\lambda}_{\mathbf{v}}} & = & \frac{1}{m}\widehat{\lambda}_{v_{\alpha}}=\widehat{\lambda}_{p_{\alpha}}.
\end{eqnarray}

\end{subequations}

The effect of $\widehat{C}$ on $m\widehat{\mathbf{V}}$ and $\widehat{\lambda}_{\mathbf{r}}$
can be computed using the well known Baker-Campbell-Hausdorff formula
$e^{X}Ye^{-X}=Y+[X,Y]+\frac{1}{2}[X,[X,Y]]+\ldots$ as follows

\begin{subequations}

\begin{eqnarray}
e^{\frac{i}{m}\mathbf{\widehat{A}}\cdot\widehat{\lambda}_{\mathbf{v}}}\left(m\widehat{V}_{\alpha}\right)e^{-\frac{i}{m}\mathbf{\widehat{A}}\cdot\widehat{\lambda}_{\mathbf{v}}} & = & m\widehat{V}_{\alpha}+i[\mathbf{\widehat{A}}\cdot\widehat{\lambda}_{\mathbf{v}},\widehat{V}_{\alpha}]
-\frac{1}{2m}[\mathbf{\widehat{A}}\cdot\widehat{\lambda}_{\mathbf{v}},[\mathbf{\widehat{A}}\cdot\widehat{\lambda}_{\mathbf{v}},\widehat{V}_{\alpha}]]+\ldots\nonumber \\
 & = & m\widehat{V}_{\alpha}+\widehat{A}_{\alpha}=\widehat{P}_{\alpha},\\
e^{\frac{i}{m}\mathbf{\widehat{A}}\cdot\widehat{\lambda}_{\mathbf{v}}}\widehat{\lambda}_{x_{\alpha}}e^{-\frac{i}{m}\mathbf{\widehat{A}}\cdot\widehat{\lambda}_{\mathbf{v}}} & = & \widehat{\lambda}_{x_{\alpha}}+\frac{i}{m}[\mathbf{\widehat{A}}\cdot\widehat{\lambda}_{\mathbf{v}},\widehat{\lambda}_{x_{\alpha}}]-\frac{1}{2m^{2}}[\mathbf{\widehat{A}}\cdot\widehat{\lambda}_{\mathbf{v}},[\mathbf{\widehat{A}}\cdot\widehat{\lambda}_{\mathbf{v}},\widehat{\lambda}_{x_{\alpha}}]]+\ldots\nonumber \\
 & = & \widehat{\lambda}_{x_{\alpha}}-\frac{1}{m}\frac{\partial\widehat{A}_{\beta}}{\partial\widehat{X}_{\alpha}}\widehat{\lambda}_{v_{\beta}}=\widehat{\lambda}_{x_{\alpha}}-\frac{\partial\widehat{A}_{\beta}}{\partial\widehat{X}_{\alpha}}\widehat{\lambda}_{p_{\alpha}}.
\end{eqnarray}

\end{subequations}

The transformation on the operators is accompanied by a transformation
on the base ket $\left|\mathbf{r},\mathbf{v}\right\rangle $ given
by

\begin{equation}
\left|\mathbf{r},\mathbf{p}\right\rangle =e^{\frac{i}{m}\mathbf{\widehat{A}}\cdot\widehat{\lambda}_{\mathbf{v}}}\left|\mathbf{r},\mathbf{v}\right\rangle =e^{\frac{i}{m}\mathbf{A}\cdot\widehat{\lambda}_{\mathbf{v}}}\left|\mathbf{r},\mathbf{v}\right\rangle =\left|\mathbf{r},\mathbf{v}-\frac{1}{m}\mathbf{A}(\mathbf{r})\right\rangle .\label{eq:legendre}
\end{equation}
By doing the unitary transformation (\ref{eq:legendre}) we are changing
the description of the state of the particle from the tangent bundle
of configuration space to the phase space.

The operators $\widehat{\mathbf{R}}$ and $\widehat{\mathbf{P}}$
act as multiplication operators on the kets $\left|\mathbf{r},\mathbf{p}\right\rangle $
defined above

\begin{subequations}

\begin{eqnarray}
\widehat{X}_{\alpha}\left|\mathbf{r},\mathbf{p}\right\rangle  & = & \widehat{X}_{\alpha}\left(e^{\frac{i}{m}\mathbf{\widehat{A}}\cdot\widehat{\lambda}_{\mathbf{v}}}\left|\mathbf{r},\mathbf{v}\right\rangle \right)=e^{\frac{i}{m}\mathbf{\widehat{A}}\cdot\widehat{\lambda}_{\mathbf{v}}}\left(\widehat{X}_{\alpha}\left|\mathbf{r},\mathbf{v}\right\rangle \right)\nonumber \\
 & = & x_{\alpha}e^{\frac{i}{m}\mathbf{\widehat{A}}\cdot\widehat{\lambda}_{\mathbf{v}}}\left|\mathbf{r},\mathbf{v}\right\rangle =x_{\alpha}\left|\mathbf{r},\mathbf{p}\right\rangle ,\\
\widehat{P}_{\alpha}\left|\mathbf{r},\mathbf{p}\right\rangle  & = & \left(m\widehat{V}_{\alpha}+\widehat{A}_{\alpha}\right)\left(e^{\frac{i}{m}\mathbf{\widehat{A}}\cdot\widehat{\lambda}_{\mathbf{v}}}\left|\mathbf{r},\mathbf{v}\right\rangle \right)\nonumber \\
 & = & e^{\frac{i}{m}\mathbf{\widehat{A}}\cdot\widehat{\lambda}_{\mathbf{v}}}\left(m\widehat{V}_{\alpha}\left|\mathbf{r},\mathbf{v}\right\rangle \right)=mv_{\alpha}\left|\mathbf{r},\mathbf{p}\right\rangle ,
\end{eqnarray}

\end{subequations} where in the last equality we used the following identity

\begin{equation}
m\widehat{V}_{\alpha}=e^{-\frac{i}{m}\mathbf{\widehat{A}}\cdot\widehat{\lambda}_{\mathbf{v}}}\left(m\widehat{V}_{\alpha}+\widehat{A}_{\alpha}\right)e^{\frac{i}{m}\mathbf{\widehat{A}}\cdot\widehat{\lambda}_{\mathbf{v}}}.
\end{equation}

On the other hand, the operators $\widehat{\lambda}'_{\mathbf{r}}$
and $\widehat{\lambda}_{\mathbf{p}}$ act as translation operators
on $\left|\mathbf{r},\mathbf{p}\right\rangle $, as it can be seen
from 

\begin{subequations}

\begin{eqnarray}
e^{-i\mathbf{a}\cdot\widehat{\lambda}'_{\mathbf{r}}}\left|\mathbf{r},\mathbf{p}\right\rangle  & = & e^{-i\mathbf{a}\cdot\widehat{\lambda}'{}_{\mathbf{r}}}e^{\frac{i}{m}\mathbf{\widehat{A}}\cdot\widehat{\lambda}_{\mathbf{v}}}\left|\mathbf{r},\mathbf{v}\right\rangle \nonumber \\
 & = & e^{\frac{i}{m}\mathbf{\widehat{A}}\cdot\widehat{\lambda}_{\mathbf{v}}}e^{-i\mathbf{a}\cdot\widehat{\lambda}{}_{\mathbf{r}}}\left|\mathbf{r},\mathbf{v}\right\rangle \nonumber \\
 & = & e^{\frac{i}{m}\mathbf{\widehat{A}}\cdot\widehat{\lambda}_{\mathbf{v}}}\left|\mathbf{r}+\mathbf{a},\mathbf{v}\right\rangle \nonumber \\
 & = & \left|\mathbf{r}+\mathbf{a},\mathbf{p}\right\rangle ,\\
e^{-i\mathbf{b}\cdot\widehat{\lambda}_{\mathbf{p}}}\left|\mathbf{r},\mathbf{p}\right\rangle  & = & e^{i\mathbf{\widehat{A}}\cdot\widehat{\lambda}_{\mathbf{v}}}\left|\mathbf{r},\mathbf{v}+\frac{\mathbf{b}}{m}\right\rangle \nonumber \\
 & = & \left|\mathbf{r},\mathbf{p}+\mathbf{b}\right\rangle .
\end{eqnarray}

\end{subequations}

In terms of the momentum operator, the Liouvillian reads

\begin{eqnarray}
\widehat{L'} & = & \frac{1}{m}\left(\widehat{P}_{\alpha}-\widehat{A}_{\alpha}\right)\widehat{\lambda}'_{x_{\alpha}}+\frac{1}{2}\left(\widehat{F}_{\alpha}\widehat{\lambda}_{p_{\alpha}}+\widehat{\lambda}_{p_{\alpha}}\widehat{F}_{\alpha}\right)\nonumber \\
 &  & +\frac{1}{2m}\frac{\partial\widehat{A}_{\beta}}{\partial\widehat{X}_{\alpha}}\left\{ \left(\widehat{P}_{\alpha}-\widehat{A}_{\alpha}\right)\widehat{\lambda}_{p_{\beta}}+\widehat{\lambda}_{p_{\beta}}\left(\widehat{P}_{\alpha}-\widehat{A}_{\alpha}\right)\right\} .\label{LP}
\end{eqnarray}

The procedure to obtain equation (\ref{LP}) explains from first principles
the origin of the minimal coupling in the KvN theory given in \cite{minimal coupling}.

We can check that the Liouvillian (\ref{LP}) is consistent with the
basic dynamic definitions (\ref{RV2}) and (\ref{LVF})

\begin{subequations}

\begin{eqnarray}
i\left[\widehat{L}',\widehat{V}_{\alpha}\right] & = & i\left[\widehat{L}',\frac{1}{m}\left(\widehat{P}_{\alpha}-\widehat{A}_{\alpha}\right)\right]=\frac{1}{m}\widehat{F}_{\alpha}(\widehat{\mathbf{R}},\widehat{\mathbf{V}}),\label{LVF-1-1}\\
i\left[\widehat{L}',\widehat{X}_{\alpha}\right] & = & \frac{1}{m}\left(\widehat{P}_{\alpha}-\widehat{A}_{\alpha}\right)=\widehat{V}_{\alpha}.\label{RV2-2}
\end{eqnarray}

\end{subequations}

The equation of motion for any state $\left|\Psi(t)\right\rangle $
is given by

\[
\frac{d}{dt}\left|\Psi(t)\right\rangle =-i\widehat{L}'\left|\Psi(t)\right\rangle .
\]

The probability $P(\mathbf{r},\mathbf{v})$ of finding the particle
at the point $(\mathbf{r},\mathbf{p})$ in the phase space is

\begin{equation}
P(\mathbf{r},\mathbf{p})=\left|\left\langle \mathbf{r},\mathbf{p}\right.\left|\Psi\right\rangle \right|^{2}.\label{bornp}
\end{equation}

The remaining elements of the Galilei algebra in terms of $\left\{ \widehat{\mathbf{R}},\widehat{\mathbf{P}},\widehat{\lambda}'_{\mathbf{r}},\widehat{\lambda}_{\mathbf{p}}\right\} $
are

\begin{eqnarray}
\widehat{\mathcal{J}'}_{\alpha} & = & \varepsilon_{\alpha\beta\gamma}\left(\widehat{X}_{\beta}\left(\widehat{\lambda}'_{x_{\gamma}}+\frac{\partial\widehat{A}_{\delta}}{\partial\widehat{X}_{\gamma}}\widehat{\lambda}_{p_{\delta}}\right)+\left(\widehat{P}_{\beta}-\widehat{A}_{\beta}\right)\widehat{\lambda}_{p_{\gamma}}\right),\label{JP}\\
\widehat{\mathcal{G}'}_{\alpha} & = & -\left(\widehat{\lambda}'_{\alpha_{\gamma}}+\frac{\partial\widehat{A}_{\delta}}{\partial\widehat{X}_{\alpha}}\widehat{\lambda}_{p_{\delta}}\right)t-m\widehat{\lambda}_{p_{\alpha}}.\label{Gp}
\end{eqnarray}

The Hilbert space spanned by the kets $\left|\mathbf{r},\mathbf{p}\right\rangle ,$the
set of operators $\left\{ \widehat{\mathbf{R}},\widehat{\mathbf{P}},\widehat{\lambda}'_{\mathbf{r}},\widehat{\lambda}_{\mathbf{p}}\right\} $,
the Liouvillian (\ref{LP}), and the elements (\ref{JP}) and (\ref{Gp})
form a unitary and irreducible, though gauge dependent, representation
of the Galilei group. This representation of the Galilei group together
with the Born rule (\ref{bornp}) is the KvN formulation of classical
mechanics\cite{cla 3}.

Let us end this section with a comparison between the wavefunctions
in the velocity vs the momentum representation. Consider the two wavefunctions
given by.
\begin{align}
\psi(\mathbf{r,v}) & =\left\langle \mathbf{r,v}\right|\left.\varPsi\right\rangle ,\nonumber \\
\varphi(\mathbf{r,p}) & =\left\langle \mathbf{r,p}\right|\left.\varPsi\right\rangle .
\end{align}

Using Eq. (\ref{eq:legendre}) we can write

\begin{align*}
\varphi(\mathbf{r,p}) & =\int\left\langle \mathbf{r,p}\right|\left.\mathbf{r}',\mathbf{v}'\right\rangle \left\langle \mathbf{r}',\mathbf{v}'\right|\left.\varPsi\right\rangle d\mathbf{r}'d\mathbf{v}'\\
 & =\int\left\langle \mathbf{r},\mathbf{v}\right|e^{-\frac{i}{m}\mathbf{\widehat{A}}\cdot\widehat{\lambda}_{\mathbf{v}}}\left|\mathbf{r}',\mathbf{v}'\right\rangle \psi(\mathbf{r}',\mathbf{v}')\,d\mathbf{r}'d\mathbf{v}'\\
 & =\int\left\langle \mathbf{r},\mathbf{v}\right|\left.\mathbf{r}',\mathbf{v}'-\frac{1}{m}\mathbf{A}(\mathbf{r}')\right\rangle \psi(\mathbf{r}',\mathbf{v}')\,d\mathbf{r}'d\mathbf{v}'\\
 & =\int\delta(\mathbf{r}'-\mathbf{r})\delta(\mathbf{v}'-\frac{1}{m}\mathbf{A}(\mathbf{r}')-\mathbf{v})\psi(\mathbf{r}',\mathbf{v}')\,d\mathbf{r}'d\mathbf{v}'\\
 & =\psi(\mathbf{r},\mathbf{v}+\frac{1}{m}\mathbf{A}).
\end{align*}

The relation between a wavefunction expressed in terms of the velocity
with the one expressed in terms of the momentum is then

\begin{equation}
\varphi(\mathbf{r,p})=\psi(\mathbf{r,v}+\frac{1}{m}\mathbf{A}).
\end{equation}

\section{Relation with Hamiltonian Mechanics}

For sake of completeness, we will show the relation between the operational
and the Hamiltonian version of classical mechanics. Since, as shown
in section 4, the operational version of classical mechanics derived
in section 3 is unitary equivalent to the KvN theory, it is enough
to show the derivation of Hamiltonian mechanics from the KvN formalism.
Let us define a wave function $\psi$ by

\begin{equation}
\psi(\mathbf{r},\mathbf{p})=\left\langle \mathbf{r},\mathbf{p}\right|\left.\psi\right\rangle .
\end{equation}

On $\psi(\mathbf{r},\mathbf{p})$ the position and momentum operators
act as multiplication operators

\begin{eqnarray}
\widehat{X}_{\alpha}\psi(\mathbf{r},\mathbf{p}) & = & x_{\alpha}\psi(\mathbf{r},\mathbf{p}),\\
\widehat{P}_{\alpha}\psi(\mathbf{r},\mathbf{p}) & = & p_{\alpha}\psi(\mathbf{r},\mathbf{p}).
\end{eqnarray}

On the other hand, the operators $\widehat{\lambda}_{\mathbf{p}}$
and $\widehat{\lambda}'_{\mathbf{r}}$ act as derivatives 

\begin{eqnarray}
\widehat{\lambda}'_{\mathbf{r}}\psi(\mathbf{r},\mathbf{p}) & = & -i\nabla_{\mathbf{r}}\psi(\mathbf{r},\mathbf{p}),\\
\widehat{\lambda}_{\mathbf{p}}\psi(\mathbf{r},\mathbf{p}) & = & -i\nabla_{\mathbf{p}}\psi(\mathbf{r},\mathbf{p}).
\end{eqnarray}

With the help of the Poisson bracket defined by 

\begin{equation}
\left\{ a,b\right\} =\frac{\partial a}{\partial x_{\alpha}}\frac{\partial b}{\partial p_{\alpha}}-\frac{\partial b}{\partial x_{\alpha}}\frac{\partial a}{\partial p_{\alpha}},
\end{equation}
we can write the components of $\widehat{\lambda}'_{\mathbf{r}}$
and $\widehat{\lambda}_{\mathbf{p}}$ as

\begin{eqnarray}
\widehat{\lambda}'_{x_{\alpha}} & = & -i\left\{ \cdot,p_{\alpha}\right\} ,\\
\widehat{\lambda}_{p_{\alpha}} & = & i\left\{ \cdot,x_{\alpha}\right\} ,
\end{eqnarray}
where the dot indicates where to put the function acted upon. For
example, for any function $f$ the operator $\widehat{\lambda}'_{x_{\alpha}}$
acts according to $\widehat{\lambda}'_{x_{\alpha}}f=-i\left\{ f,p_{\alpha}\right\} $
.

The Liouvillian operator (\ref{LP}) can be written as

\begin{equation}
\widehat{L}=-i\left\{ \cdot,H\right\} ,
\end{equation}
where $H$ is the classical Hamiltonian given by

\[
H=\frac{\left(\mathbf{p}-\mathbf{A}(\mathbf{r})\right)^{2}}{2m}+V(r).
\]

The Schr\"odinger-like equation (\ref{defL2}) becomes

\begin{equation}
\frac{\partial\psi}{\partial t}+\left\{ \psi,H\right\} =0,\label{Liouvillepsi}
\end{equation}
or written for the complex conjugate $\psi^{*}$

\begin{equation}
\frac{\partial\psi^{*}}{\partial t}+\left\{ \psi^{*},H\right\} =0.\label{Liouvillepsi2}
\end{equation}

Equations (\ref{Liouvillepsi}) and (\ref{Liouvillepsi2}) are linear
in the derivatives and they can be combined into the single equation

\begin{equation}
\frac{\partial\rho}{\partial t}+\left\{ \rho,H\right\} =0,\label{cliouville}
\end{equation}

where $\rho=\left|\psi\right|^{2}$. Equation (\ref{cliouville})
is the classical Liouville equation for the evolution of the probability
distribution in phase space. Thus, the operational formulation of
classical mechanics is equivalent to classical statistical mechanics.

Let us note that the following probability density, known as the Klimontovich
distribution for the particle \cite{ichimatsu}, given by

\begin{equation}
\rho(t)=\delta(\mathbf{r}-\mathbf{r}(t))\delta(\mathbf{p}-\mathbf{p}(t)),\label{kilmontovich:}
\end{equation}
is a solution to Liouville's equation (\ref{cliouville}) as long
as $x_{\alpha}(t)$ and $p_{\alpha}(t)$ are solution to Hamilton
equations

\begin{eqnarray}
\frac{dx_{\alpha}}{dt} & = & \frac{\partial H}{\partial p_{\alpha}},\\
\frac{dp_{\alpha}}{dt} & = & -\frac{\partial H}{\partial x_{\alpha}}.
\end{eqnarray}
The probability distribution (\ref{kilmontovich:}) represents a point
in phase space and that point will move along a trajectory given by
the solution of the Hamilton equations. 

\section{Comparison with the Quantum Case}

For the sake of comparing the classical and quantum unitary representations
of the Galilei group, let us restate the commutation relations of
the algebra in terms of the usual symbols used in quantum mechanics
$(\widehat{\mathbf{J}},\,\widehat{\mathbf{P}},\,\widehat{\mathbf{G}},\,\widehat{H},\,M)$.
The Galilei algebra for the vanishing commutators read 

\begin{equation}
\left[\widehat{P}_{\alpha},\widehat{P}_{\beta}\right]=\left[\widehat{G}_{\alpha},\widehat{G}_{\beta}\right]=\left[\widehat{J}_{\alpha},\widehat{H}\right]=\left[\widehat{P}_{\alpha},\widehat{H}\right]=0,\label{QG 1}
\end{equation}
while the non-vanishing ones are
\begin{align}
\left[\widehat{J}_{\alpha},\widehat{J}_{\beta}\right] & =i\varepsilon_{\alpha\beta\gamma}\widehat{J}_{\gamma};\;\left[\widehat{J}_{\alpha},\widehat{P}_{\beta}\right]=i\varepsilon_{\alpha\beta\gamma}\widehat{P}_{\gamma};\nonumber \\
\left[\widehat{J}_{\alpha},\widehat{G}_{\beta}\right] & =i\varepsilon_{\alpha\beta\gamma}\widehat{G}_{\gamma};\;\left[\widehat{P}_{\alpha},\widehat{G}_{\beta}\right]=i\delta_{\alpha\beta}M;\nonumber \\
\left[\widehat{G}_{\alpha},\widehat{H}\right] & =i\widehat{P}_{\alpha}.\label{QG2}
\end{align}

As was done for the classical case, the idea is to look for a Hilbert
space $\mathcal{H}_{Q}$ where the space-time transformation of the
Galilei group are represented by

\begin{eqnarray*}
\mathrm{\mathbf{Space-Time\:Transformations}} &  & \mathbf{Unitary\:Operator}\\
\mathrm{Rotations} & \qquad & \mathrm{}\\
\mathbf{x}\rightarrow R_{\alpha}(\theta_{\alpha})\mathbf{x} & \qquad & e^{-i\theta_{\alpha}\widehat{J}_{\alpha}}.\\
\mathrm{Spatial\:Displacement}\\
\mathbf{x}\rightarrow\mathbf{x}+\mathbf{a} &  & e^{-ia_{\alpha}\widehat{P}_{\alpha}}.\\
\mathrm{Galilean\:Boost}\\
\mathbf{x}\rightarrow\mathbf{x}+\mathbf{b}t &  & e^{ib_{\alpha}\widehat{G}_{\alpha}}.\\
\mathrm{Time\:Displacement}\\
t\longrightarrow t+\tau &  & e^{i\tau\widehat{H}}.
\end{eqnarray*}

Ignoring the spin degrees of freedom, the components of the position
operator $\widehat{\mathbf{R}}$ are assumed to form a complete set
of commuting observables and the Hilbert space $\mathcal{H}_{Q}$
is build using kets of the form $\left|\mathbf{r}\right\rangle $.
With this choice of Hilbert space, and assuming relation (\ref{RV1})
for the position and velocity, the usual relations for quantum mechanics
for the particle can be found \cite{jordan,ballentine}

\begin{align}
\left[\widehat{X}_{\alpha},\widehat{P}_{\beta}\right] & =i\delta_{\alpha\beta},\\
\widehat{\mathbf{J}} & =\widehat{\mathbf{R}}\times\widehat{\mathbf{P}},\\
\widehat{\mathbf{G}} & =M\widehat{\mathbf{R}}-t\widehat{\mathbf{P}},\\
\widehat{H} & =\frac{P^{2}}{2M}.
\end{align}

A difference of the quantum representation over the classical one
is that all the operators involved in the algebra have direct physical
significance: $M$ is the mass, $\widehat{\mathbf{P}}$ is the momentum,
$\widehat{\mathbf{J}}$ is the angular momentum, $\widehat{H}$ is
the energy and $\widehat{\mathbf{G}}$ is a physical quantity sometimes
called the dynamic mass moment. The above is in contrast to the classical
case where it is hard to say what is the physical significance of
$\widehat{L}$ or $\widehat{\lambda}_{x_{\beta}}$\cite{cla 3}. On
the other hand, due to the non-vanishing of $M$ (quantum representation
with vanishing $M$ are unphysical \cite{levy}) the quantum case
end up being a projective (or ``up to a phase'') representation.
In the classical case, as we shown in section 3, the central charge
$\mathcal{M}$ is not only allowed to vanish but it is actually required
to do so.

The origin of the difference between the two theories lies in the
amount of information than a state can contain in relation to the algebra of operators that act on the Hilbert space \footnote{The difference can not be in the mathematical structure of the Hilbert space used because both classical and quantum mechanical Hilbert spaces are isomorphic to the space of square-integrable functions.}. In quantum mechanics the number of mutually
commuting operators is restricted to three compared with the classical case where
complete set consist of six operators, three from the position and
three from the momentum (or velocity).

From the point of view of ``wave mechanics'', the choice of the different algebra of operators acting on the Hilbert space now means that the classical $\psi$ and $\rho$
are functions of phase space (or the tangent bundle of configuration space), unlike the quantum case where $\psi$
and $\rho$ are functions  of configuration space alone. The structure
of the two theories is markedly different. As an example, we have
seen that the classical Schr\"odinger-like equation in phase space is
the Liouville equation (\ref{cliouville}), a equation of first derivatives
in contrast with the second derivatives appearing in the Schr\"odinger
equation. As a result in quantum mechanics the wave functions squared
is the probability density in volume whereas in the classical case it is the
probability density in phase space.
 Moreover, going to the polar representation 

\begin{equation}
\psi=\sqrt{\rho}e^{iS},\label{polar}
\end{equation}
the two theories differ in the sense that by replacing (\ref{polar}) in the Schr\"odinger equation a set of two coupled differential equations is obtained
giving rise to the Madelung fluid \cite{madelung} which sparked a possible new interpretation of quantum mechanics (see, e.g., \cite{wyatt}). For the classical case, after replacing (\ref{polar}) into the Liouville equation the resulting equations for $\rho$ and $S$ are decoupled \footnote{It has been shown that in the classical case
S can be related to the classical action \cite{cla 6}}.

Finally, let us mention another aspect of the operational version of classical mechanics that is different from the quantum case, the superposition of states. Coherent superposition of states has been studied in the context of the KvN theory \cite{Cla 8,gozzi}(the same analysis holds if we use the velocity operator instead the momentum) and there seems to be two possibilities. (1) If the observables of the theory are only the operators corresponding to phase space variables (position, momentum and any combination of them) then a superselection rule gets triggered, coherent superposition cannot be distinguished from a mixed density matrix and relative phases cannot be detected. This is somewhat strange compared to the quantum case, but it makes some sense if we consider that the pure state of classical mechanics are delta functions in phase space and all other densities are statistical states. (2) To avoid the superselection mechanism the auxiliary operators have to be included as observables. As a consequence, this will make the phases in the KvN wavefunctions to be observable also. It is known, for example, that the lack of degeneracy of the zero eigenvalue of the Liouvillian implies ergodicity \cite{arnold}, so it is tempting to prefer this posibility. However, since we are increasing the number of observables, this path seems to imply a more general theory that just classical mechanics. In the end, the question of which of the above two choices is the better seems to be unanswered\footnote{ A clue to this question might be obtained from a measurement theory of the KvN formalism \cite{so} that includes analysis of operators like the Liouvillian}.

\section{Discussion and Concluding Remarks}

The operational formulation of classical mechanics given in the article
is completely independent from any quantum result. The equations given
in the present work were not derived as a classical limit of some
quantum equations. Just as the non-relativistic quantum mechanics
can be derived, under certain assumptions, from a unitary representation
of the Galilei group, we have shown that a Hilbert space formulation
of classical mechanics (leading later to the Koopman-von Neumann theory)
can be obtained as a unitary representation of the same group. The
derivations, albeit similar in spirit, differ in a crucial assumption.
In the classical case we demand the base kets $\left|\mathbf{r},\mathbf{v}\right\rangle $
to be eigenvectors of the position and the velocity operators simultaneously,
in contrast with the quantum mechanical case

Our formalism stresses the fact that in classical mechanics, and in
contrast with quantum mechanics, the potentials are auxiliary optional
quantities. We can see from equations (\ref{defL2}) and (\ref{LVF-1})
that the force operator is sufficient to give a complete description
of the evolution of the state of the classical particle. Moreover,
the formalism allows forces that cannot be derived from any generalized
potential. 

There are several aspects that can be analyzed in further investigations.
For example, through the article we only used Cartesian coordinates
but we know classical mechanics gives plenty of freedom to choose
other charts. Hence, the following question arises: what is the role
of constraints and generalized coordinates in the operational version
of classical mechanics?

Another relevant result that requires further clarification is the
relation between the quantum canonical transformation we used to go
from the irreducible set of operators $\left\{ \widehat{\mathbf{R}},\widehat{\mathbf{V}},\widehat{\lambda}_{\mathbf{r}},\widehat{\lambda}_{\mathbf{v}}\right\} $
to the set $\left\{ \widehat{\mathbf{R}},\widehat{\mathbf{P}},\widehat{\lambda}'_{\mathbf{r}},\widehat{\lambda}_{\mathbf{p}}\right\} $
and the Legendre transformation used to pass from Lagrangian mechanics
to Hamiltonian mechanics. From a more abstract point of view, the
rays $\left\{ \left|\mathbf{r},\mathbf{v}\right\rangle \right\} $
and $\left\{ \left|\mathbf{r},\mathbf{p}\right\rangle \right\} $
are in a one to one correspondence with points in the tangent and
the cotangent bundles of configuration space, respectively. We demonstrated
that the set of kets $\left|\mathbf{r},\mathbf{v}\right\rangle $
and $\left|\mathbf{r},\mathbf{p}\right\rangle $ are related by a
an unitary transformation. It remains to be seen
what is the exact relation between our treatment and the geometrical
one of analytical mechanics.

Besides the purely classical results, we hope that our work here helps in the advancement of interesting research topics that probe
into the interface between quantum mechanics and the classical physics, such like
 decoherence, quantum chaos,  the hydrodynamical interpretation
of quantum mechanics, or quantum-classical hybrid systems \cite{Hybryd,Hybryd2}. With a Hilbert space at hand closely related to classical mechanics such
research are enriched with new perspectives. In this paper we laid a solid ground for this kind
of investigations by showing that both quantum and classical mechanics can be based on unitary representation of the Galilei group.

\end{document}